\documentclass[12pt]{article}
\usepackage{epsfig}
\usepackage{graphicx}
\usepackage{rotate}
\usepackage{epsf}
\def\speaker#1{{\bf #1:}\ }

\newcommand{\Alfven}{$\rm Alfv\acute{e}n$}

\newcommand{\DD}{\frac}

\newcommand{\ber}{\begin{array}}
\newcommand{\eer}{\end{array}}

\newcommand{\lra}{\longrightarrow}

\newcommand{\Msun}{{\,{\cal M}_{\odot}}}

\newcommand{\Mdot}{{\,\dot{\cal M}}}
\def\beq{\begin{equation}}
\def\eeq#1{\label{#1}\end{equation}}
\def\eeqn{\end{equation}}

\def\beqa{\begin{eqnarray}}
\def\eeqa#1{\label{#1}\end{eqnarray}}
\def\eeqan{\end{eqnarray}}

\let\bar=\overbar

\def\Dslash{\not{\hbox{\kern-4pt $D$}}}
\def\dslash{\not{\hbox{\kern-2pt $\del$}}}

\def\msb{{\bar{\ssstyle M \kern -1pt S}}}

\def\Title#1{\begin{center} {\Large {\bf #1} } \end{center}}

\begin{document}

\Title{Jet-formation in black holes accreting systems  and in GRBs }

\bigskip\bigskip


\begin{raggedright}  

{\it Ahmad Hujeirat  \index{Reggiano, D.}\\
 Max-Planck-Institut f\"ur Astronomie\\
D-69117 Heidelberg, Germany }
\bigskip\bigskip
\end{raggedright}

\section{Abstract}
Most powerful jets are observed to emanate from accreting black hole systems.
              Recent quasi-stationary radiative MHD calculations reveal that
              jet-launching start in the innermost region of a 
              transition layer (-TL), which is located between
              the disk and the corona.
              The plasma in the TL is dominated by virial-hot protons,  
              advective, highly magnetic-diffusive and rotates super-Keplerian.
              The amplified toroidal magnetic field (TMF) in the TL reach values beyond
              equipartition with respect to the electron thermal energy, thereby 
              considerably enhancing the Poynting energy flux and 
              yielding gravitationally unbound electron-proton outflows with a
              positive Bernoulli number.  

              We speculate that beaming jets in GRBs may possess similar properties, 
              provided that the GRB-progenitors have rotational 
              energy sufficient for forming a nearly Keplerian and massive disk.

\section{Introduction}

 Several observational signatures have been discovered in the 
last years suggesting that the central nucleus might be directly involved
in the initiation and acceleration of the powerful jets observed in
AGNs and BHXBs \cite{Mirabel}. 
In particular, Kerr black holes surrounded by magnetized accretion disks
are considered to be central engines that power the observed superluminal
jets (e.g., M87 and GRS 1915+105, see \cite{Rees}, \cite{Blandford}).
Those spinning black holes accreting at extremely low rates are expected to power 
electron-positron jets, whilst electron-proton jets emerge from BHs accreting  at higher
accretion rates. In the case of accretion at super-Eddington rates, the associated
energies is likely to be trapped in the accreted matter, and subsequently affect the 
mass-growth of the central engine directly. The later case is less
studied, and therefore it is not clear what and how the back reaction of the BH would be.

On the other hand, recent observations reveals that GRBs might be located in
star-forming regions, whilst their long time duration rises  
the possibility that they are violent ends of massive star evolution 
\cite{Paczynski}, \cite{Rhoads}.
Recent hydrodynamical studies of star formation, however, reveal that 
the first stars in the universe must have been massive, slowly rotating and weakly
magnetized see \cite{Bromm}, \cite{Abel}. Such population of massive stars,
the so called population III, are considered to be unstable, and they undergo 
dynamical collapse 
to form massive black holes \cite{Woosley}, \cite{Piran}. 
Here the iron core of the massive star collapses to form a black hole surrounded by
an accretion disk.
Magnetic fields and neutrinos in this model are essential components for extracting
rotational and gravitational energies to power relativistic jets along the
polar axis \cite{MacFadyen}.

In the present paper, we address the possibility that, prior to  GRB-events, 
the configuration is similar to that of a BH surrounded by a massive disk, and
accreting at extremely large super-Eddington rates.
\section{The role of magnetic fields}
In the outermost regions of a disk, we anticipate the magnetic fields to be
below equipartition. 
Let $r_\mathrm{tr}$ be a transition radius (see Fig. 1), such that for 
$r > r_\mathrm{tr}$, we have
the usual standard disk \cite{Shakura}, where the ratio of magnetic to gas pressure 
$\beta= P_\mathrm{mag}/P_\mathrm{gas} \ll 1$. $r \le r_\mathrm{tr}$ is the 
region where magnetic fields (-MFs) are in equipartiation with thermal energy of the plasma,
or even stronger. In the region $r > r_\mathrm{tr}$, the Balbus-Hawley instability 
\cite{Balbus}
operates on the dynamical time scale: it rapidly amplifies the MF within
 this time scale and forces
$\beta$ to approach, but remains below, unity \cite{Hawley}.
Thus, when the matter at the outer region has completed one revolution around
the central BH, MFs should have reached equipartition almost everywhere in the disk.
These strong MFs will eventually suppress the generation of turbulence, and so
terminating angular momentum transport through any sort of friction.

 As the rotational energy in classical Keplerian disks exceeds the
thermal energy by at least two orders of magnitude, the generated magnetic energy 
via shear (mainly toroidal) can easily become beyond equipartition with respect 
to the internal energy.\\ \\
How does accretion proceed under $\beta = 1$ conditions?\\ \\
In general, accretion via strongly magnetized disks proceeds if the MFs can extract
angular momentum. The efficiency of extraction depends strongly on the MF-topology. 
For example, large scale or dipolar MF-topologies are considered to be appropriate
for extracting rotational energy from the disk  or even from the hole itself
and powers the jets.

We may use the equations of mass and magnetic flux conservation to 
estimate the radial variation of the MFs in the vicinity of the BH. The 
equation of mass conservation in 1D reads: 
\begin{equation}
  \DD{\partial \Sigma}{\partial t} + \DD{1}{r}\DD{\partial}{\partial r} r \Sigma U = 0, \\
\end{equation}
where $\Sigma$ is the  surface density. 
In the steady state case, this is equivalent to: 
\begin{equation}
  r <\Sigma U^-> - r <\Sigma U^+> \approx  r <\Sigma U^-> = const.,
\end{equation}
where the superscripts $-,\,+$ resemble the mean value of the in- and 
out-flowing plasmas, respectively.
The flux $ r <\Sigma U^+> $ can be safely neglected, as most observed
jets are expected to carry a negligibly small fraction of the matter in the disk.
In this case, the surface density $\Sigma$ may adopt one of the following profiles:
\begin{equation}
 \Sigma \sim  \left \{ 
     \begin{array}{ll}
      r^{-3/4} &  {\rm \hspace*{1.0cm} standard\, disks} \\
      r^{-2/3} &  {\rm \hspace*{1.0cm} advective\, disks}\footnote{} \\
      r^{-1/2} &  {\rm \hspace*{1.0cm} ADAF-disks}\footnote.
    \end{array} \right.
\end{equation}
 \footnotetext[1]{ These solutions correspond to disk-models in which 
               $ U \approx V_\mathrm{A} \approx V_\mathrm{s} \sim r^{-1/3}$
                and  $\Omega $ is roughly Keplerian \cite{Hujeirat3}. $V_\mathrm{s}$ here is
                the sound speed.}
 \footnotetext[2]{See \cite{Narayan}.
                   In writing  $r^{-1/2}$ we have taken into account that 
                     the vertical length scale height scales linearly with radius}
Taking into account the global conservation of the magnetic flux normal to the disk, i.e.,
  $\Phi = 2 \pi r^2 B_\mathrm{p} \approx 2 \pi r^2 B_\mathrm{\theta} = const.$,
one can easily verify that:

\begin{equation}
 B_\mathrm{p} \sim  \left \{ 
     \begin{array}{ll}
      \Sigma^{8/3}   &  {\rm \hspace*{1.0cm} standard\, disks} \\
      \Sigma^{3} &  {\rm \hspace*{1.0cm} advective\, disks} \\
      \Sigma^{4}   &  {\rm \hspace*{1.0cm} ADAF-disks.}
    \end{array} \right.
\end{equation}

Consequently, if the disk extends down to the last stable orbit, magnetic fields
will increase rapidly inwards, diminishing turbulence-generation and making torsional
{\Alfven}  waves as the unigue carrier of angular momentum in the vicinity of the BH.

Therefore, unless the imposed boundary conditions limit the amplification of TMF,
there is no reason, why $\beta$ should remain smaller than unity.  

Another conclusion which can be drawn from the strong dependence of $B_\mathrm{p}$
on $\Sigma$, irrespective of the underlying model, is that the disk must truncate near
the last stable orbit and become a TDAT-disk \cite{Hujeirat1}, otherwise
a magnetic barrier will be formed that terminate accretion.
 
To clarify the action of magnetic braking, we write the conservation equation
of angular momentum $\ell (\doteq \rho r^2 \cos{\theta}\, \Omega) $ as follows:
\begin{figure}[htb]
\begin{center}
{\hspace*{-0.5cm}
\includegraphics*[width=7.5cm, bb=1 1 350 256,clip]
{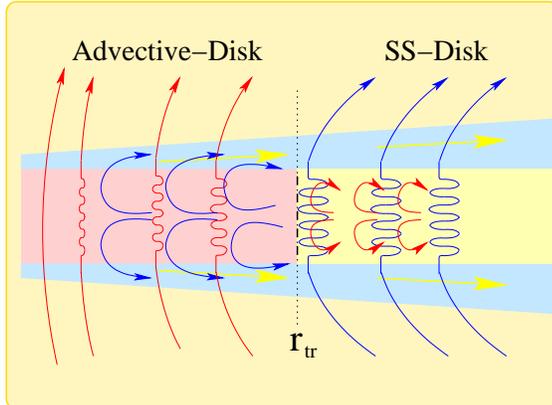}
}
\end{center}
{\vspace*{-0.4cm}}
\caption [ ] { The transition from a standard to an advection-dominated disk.
The poloidal MF (-PMF) in the outer regions is weak and below equipartition, and 
hydrodynamical turbulence is expected to be the dominant mechanism for angular
momentum transport. This is however a transient case, as the BH-instability
will amplify the MF up to equipartition on approximately the dynamical time scale.
 In the innermost region of the disk, 
 MFs  are in equipartition with the thermal energy, hence diminishing the 
generation of turbulence. In this region, transport of angular momentum is mediated mainly
by torsional {\Alfven}  waves. 
 At $R_\mathrm{tr}$ we expect the equality T1=T3 (see Eq. 5), i.e., the tranport
angular momentum via hydrodynamical turbulence is as efficient as the
vertical transport via MFs.      
  } 
\end{figure}
\begin{figure}[htb]
\begin{center}
{\hspace*{-0.5cm}
\includegraphics*[width=6.5cm, bb=0 0 270 240,clip]
{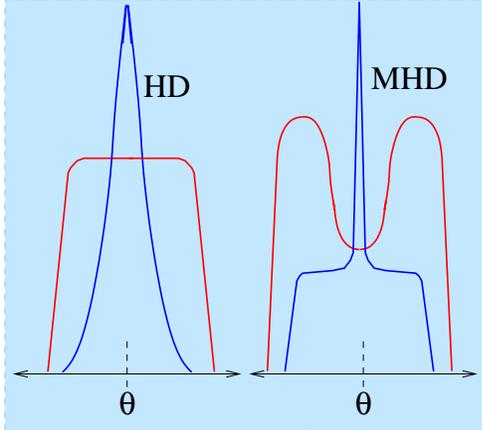}
}
\end{center}
{\vspace*{-0.4cm}}
\caption [ ] { The vertical distribution of rotational velocity
               in the disk with and without MFs. The action of the
  equipartition MFs is to transport angular momentum from the
  equatorial near-regions to higher latitudes, where the particles 
  experience  centrifugal acceleration and start to move outwards.
  } 
\end{figure}
\begin{equation}
 \DD{\partial \ell }{\partial t}  + \overbrace{\nabla\cdot V \ell}^{\rm T0}
= \overbrace{B_\mathrm{r} \DD{\partial B_{\rm T}}{\partial \mathrm{r}}}^{\rm T1}
  +  \overbrace{B_\theta \DD{\partial B_T}{\partial \theta}}^{\rm T2}
  +  \overbrace{
 \DD{1}{r^2} \DD{\partial}{\partial r}  r^4 \rho \nu \DD{\partial \Omega}{\partial r}}^{\rm T3}
 + \overbrace{
 \DD{1}{\cos{\theta}} \DD{\partial}{\partial \theta}  \cos{\theta} \rho \nu
 \DD{\partial \Omega}{\partial \theta}}^{\rm T4}.
\end{equation}
 
T0 denotes angular momentum transport via advection, T1 and T2 are for magnetic
extraction, and T3 and T4 are for viscous
(microscopic or hydrodynamical turbulent) re-distribution of the angular
momentum. In classical accretion disks, we have the equality T0=T3, whilst all other 
terms are neglected,  traditionally. In general,
the term T1 can be safely neglected in the equatorial near-plane if the 
MFs have a large or dipolar topology. Although it is more reasonable to consider
T1 as the dominant term on the RHS of Eq. 5 than T3, non of the above terms
 will be neglected in the present study. 

Combining Eq.5 with the equation corresponding to the time-evolution of the
TMF, and assuming the flow to be nearly incompressible, we obtain the
torsional wave equation which has the approximate form: 

\begin{equation}
 \DD{\partial^2 B_\mathrm{T} }{\partial^2 t}  = {V^2_\mathrm{A}}   
       \Delta {B_{\rm T}}.
\end{equation}
$\Delta$ denotes the two-dimensional Poisson operator in spherical geometry, and 
$V_\mathrm{A}$ ($\doteq B_\mathrm{p}/\sqrt{\rho}$)  is the {\Alfven} speed.
The action of these waves is to magnetically brake the disk through transporting angular 
momentum from the equatorial near-plane to higher latitudes. Note that $B_\mathrm{p}$
uniquely determines the speed of propagation, hence the dependence of the efficiency of
angular momentum transport on the $B_\mathrm{p}-$topology.
For a nearly  Keplerian-rotating medium, the removal time scale of angular momentum is: 
 \begin{equation}
  \tau_\mathrm{rem} \sim \rho V_\mathrm{T} H/B_\mathrm{P} B_\mathrm{T} \sim r^{3/2}.
\end{equation}
This implies, for example, that the rate at which angular momentum is removed
at the last stable orbit ($\doteq R_\mathrm{LSO}$) is one thousand times faster
than at  $r=100 R_\mathrm{LSO}$. To avoid the innermost region
of the disk from running out of angular momentum, we require the disk to be dynamically stable.
Equivalently,  the rate at which angular momentum is removed at any radius  must be 
equal to the rate at which it is advected
inwards from the outer layers, i.e., $\tau_\mathrm{adv}= \tau_\mathrm{rem}$. 
This implies that the radial velocity
$-U$  in the disk is of the order of the {Alfven} speed, which is  also of the order
of the sound speed $V_\mathrm{S}$, hence the terminology advection-dominated disk. 
 
\begin{figure*}[htb]
\begin{center}
{\hspace*{-0.5cm}
\includegraphics*[width=12.0cm, bb=60 393 511 739,clip]
{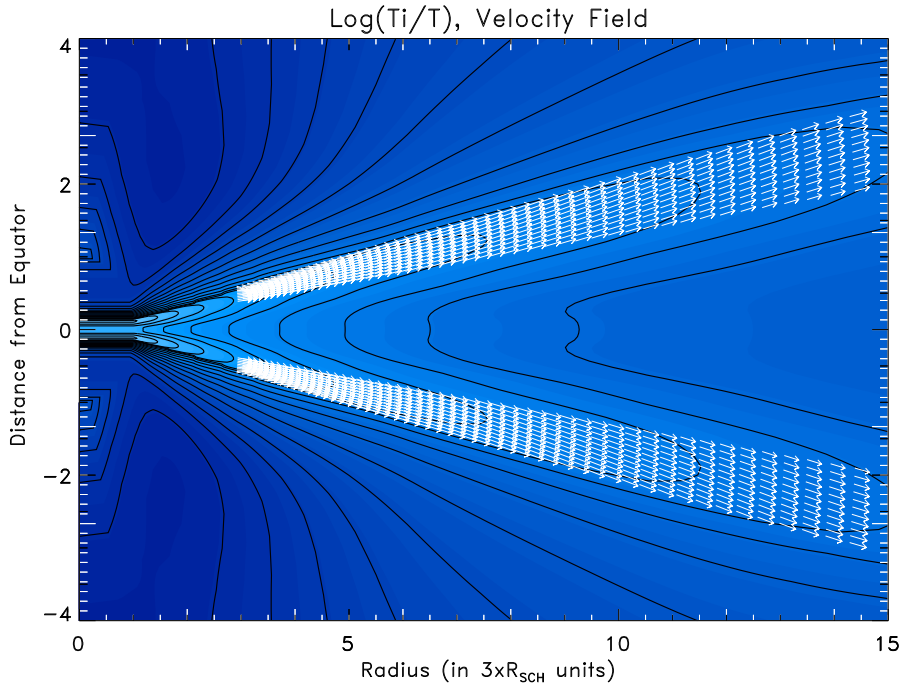}
}
\end{center}
{\vspace*{-0.4cm}}
\caption [ ] { The distribution of the velocity field in the transition region 
           superposed on equally-spaced isolines of the
          angular velocity $\Omega$. The strong-decrease of $\Omega$
          with radius in the equatorial region relative to its slow-decrease in the
          TL is obvious.  
  } 
\end{figure*}

\section{Accelerating the magnetized plasma into jets}

When an isolated Keplerian-rotating particle in the equatorial near-plane is shifted to
higher latitudes while conserving its angular momentum, its rotation becomes super-Keplerian
locally, and therefore it starts to move outwards. Moreover, noting that $\tau_\mathrm{rem}$ and $H_\mathrm{disk}$
decrease with decreasing radius, more particles in the innermost region of the disk  
are expected to be shifted to higher latitudes than in the outer regions, provided that
 $\beta \approx 1$ and $T_\mathrm{i} \approx T_\mathrm{virial}$. This is because
 particles prefer to move along MF-lines than across them.  As a consequence,
a transition layer between the disk and the hot tenuous corona starts to evolve, 
where the matter is enforced to  
rotate super-Keplerian, and start to accelerate outwards.
This configuration places an important role for the central engine as a source for 
powering jets.
Based on a previous calculations \cite{Hujeirat2}, 
it was found that that   1)  the angular velocity  
in the transition layer (TL) adopts approximately the super-Keplerian profile
$\Omega \sim r^{-5/4}$. 2) Energy dissipation is injected primarily 
into the ions or protons that cool predominantly through fast outflows. 
3) The amplification of the generated toroidal magnetic field is 
   terminated by the magnetic diffusion (reconnection) and fast outflows. 
The width of the TL is pre-dominantly determined through the transverse
variation of the ion-pressure across the jet, i.e.,  $ H_\mathrm{W} = 
P_\mathrm{i}/\nabla P_\mathrm{i} \approx 0.2\,r$,
and so strongly dependent on whether the flow is a one- or two-temperature plasma.
In the steady-state case, this implies:
\begin{equation}
   \rho   \sim r^{-7/4},
   T_\mathrm{i}  \sim r^{-1/2},
   U_\mathrm{r}  \sim r^{-1/4},
   B_\mathrm{P}/B_\mathrm{T} \sim \rm const.
\end{equation}

We note that  $U_\mathrm{r}$ adopts a profile and attains values similar  to those 
in the innermost part of the disk.
Provided that energy exchange between the matter
in the disk and in the TL is efficient, the incoming matter can easily  be
re-directed into outwards-oriented motions. This implies that the Bernoulli number (Be) 
can change sign in dissipative flows. As Fig. 4 shows, 
Be is  everywhere negative save the TL, where it attains large positive
values, so that the ion-plasma can start its kpc-journey.
We note that the outflow is sufficiently strong to shift the PMF lines outwards, 
whilst the large magnetic diffusivity prevents the formation of large electric currents
along the equator. In the corona
however, MFs are too weak to halt the diffusive plasma in the dynamically unstable
corona against gravity, and instead, they drift with the
infalling gas inwards. In the case of very weak MFs ($\beta \le 0.1$),  
our calculations indicate a considerably weak outflow.  This is a consequence of the tendency of
the MFs to establish a monopole like-topology, i.e, a one-dimensional MF-topology in which 
$\rm{B}_\theta \lra 0$.
In this case, the magnetic tension  becomes inefficient in feeding  the matter
in the TL with the angular momentum required for launching jets, indicating herewith that
cold accretion disks alone are in-appropriate for initiating winds (see \cite{Ogilvie}).

Comparing the flux of matter in the wind to that in the disk, it has been found that
${\Mdot}_\mathrm{W}/{\Mdot}_\mathrm{d} = {\rm const.} \approx 1/20$. The angular momentum flux associated with
the wind  is
 $\dot{\cal J}_\mathrm{W}/\dot{\cal J}_\mathrm{d} = {\rm a }\, 
({\Mdot}_\mathrm{W}/{\Mdot}_\mathrm{d})\,r^{1/4},$ where ${\rm a }$ is a constant of order unity.
Consequently, at 500 gravitational radii almost $25\%$ of the total accreted angular momentum
in the disk re-appears in the wind. 

\begin{figure*}[htb]
\begin{center}
{\hspace*{-0.5cm}
\includegraphics*[width=12.5cm, bb=55 390 500 740,clip]
{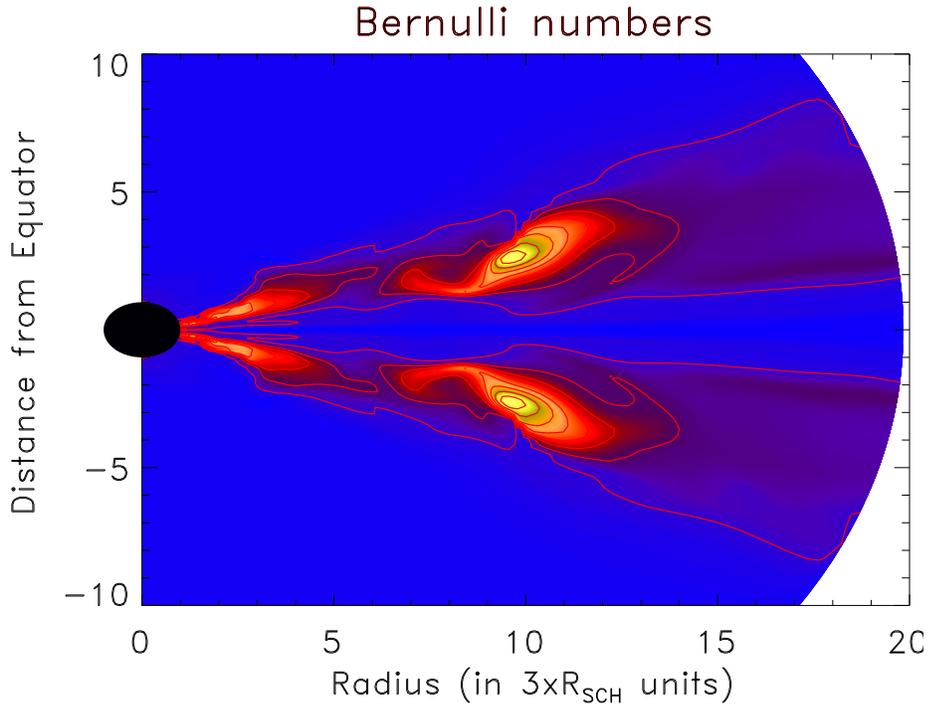}
}
\end{center}
{\vspace*{-0.4cm}}
\caption [ ] { A snap-shot of the distribution of the Bernoulli number in 
               two-dimensional quasi-stationary calculations. 
              The decrease from large to low positive values is represented
              via yellow, green and red colors.  The blue color corresponds
              to negative values.  The figure shows two ejected gravitationally unbound 
              blobs of large positive energies in the TL. Ejection starts from the
             innermost region of the TL and evolves non-linearly.  } 
\end{figure*}
 

To clarify why the TL is geometrically thin, we note that 
in stratified dissipative flows the density scale height is much smaller than 
the scale height of the angular velocity (see \cite{Hujeirat2}). 
Since the flow in TL is highly dissipative and rotates super-Keplerian, the time
scale for torsional {Alfven} waves to cross the layer vertically is longer than the
dynamical time scale at which angular momentum is advected outwards. 

On the other hand, unless there is a significant energy flux that heats up the plasma 
from below, as in the case of stars, heat conduction will always force the BH-coronae
to collapse dynamically.  
To elaborate this point, let us compare the conduction time scale with the dynamical
time scale along $\rm B_\mathrm{P}$-field at the last stable orbit of a 
SMBH:
\[
\DD{\tau_{\rm cond}}{\tau_{\rm dyn}} =  \DD{r \rho U_\mathrm{r}}{\kappa_0 T^{5/2}_\mathrm{i}}
 = 4.78\times 10^{-4} \rho_{10} T^{-5/2}_\mathrm{i,10} {\cal M}_8, \]
where $\rho_{10}$, $\rm{T_\mathrm{i,10}}$ and ${\cal M}_8$ 
 are respectively in $10^{-10}\,\rm{g\,cm^{-3}}$, 
 $10^{10}\,$K and in $10^8\,\Msun$ units.
This is much less than unity for most reasonable values of density and temperature typical for 
AGN-environments. In writing Eq. 4 we have taken optimistically the upper limit $ c/\sqrt{3}$ for the velocity,
and set $\kappa_0 = 3.2\times 10^{-8}$ for the ion-conduction coefficient.
When modifying the conduction operator to respect causality, we obtain 
${\tau_{\rm cond}}/{\tau_{\rm dyn}} \le U_\mathrm{r}/c $, which is again smaller than unity.
 \\
This agrees with our numerical calculations which rule out the 
possibility of outflows  along the rotation axis, and in particular not 
from the highly unstable polar region of the BH, as ADAF-solutions predict.
\begin{figure}[htb]
\begin{center}
{\hspace*{-0.5cm}
\includegraphics*[width=10.5cm, bb=46 270 375 505,clip]
{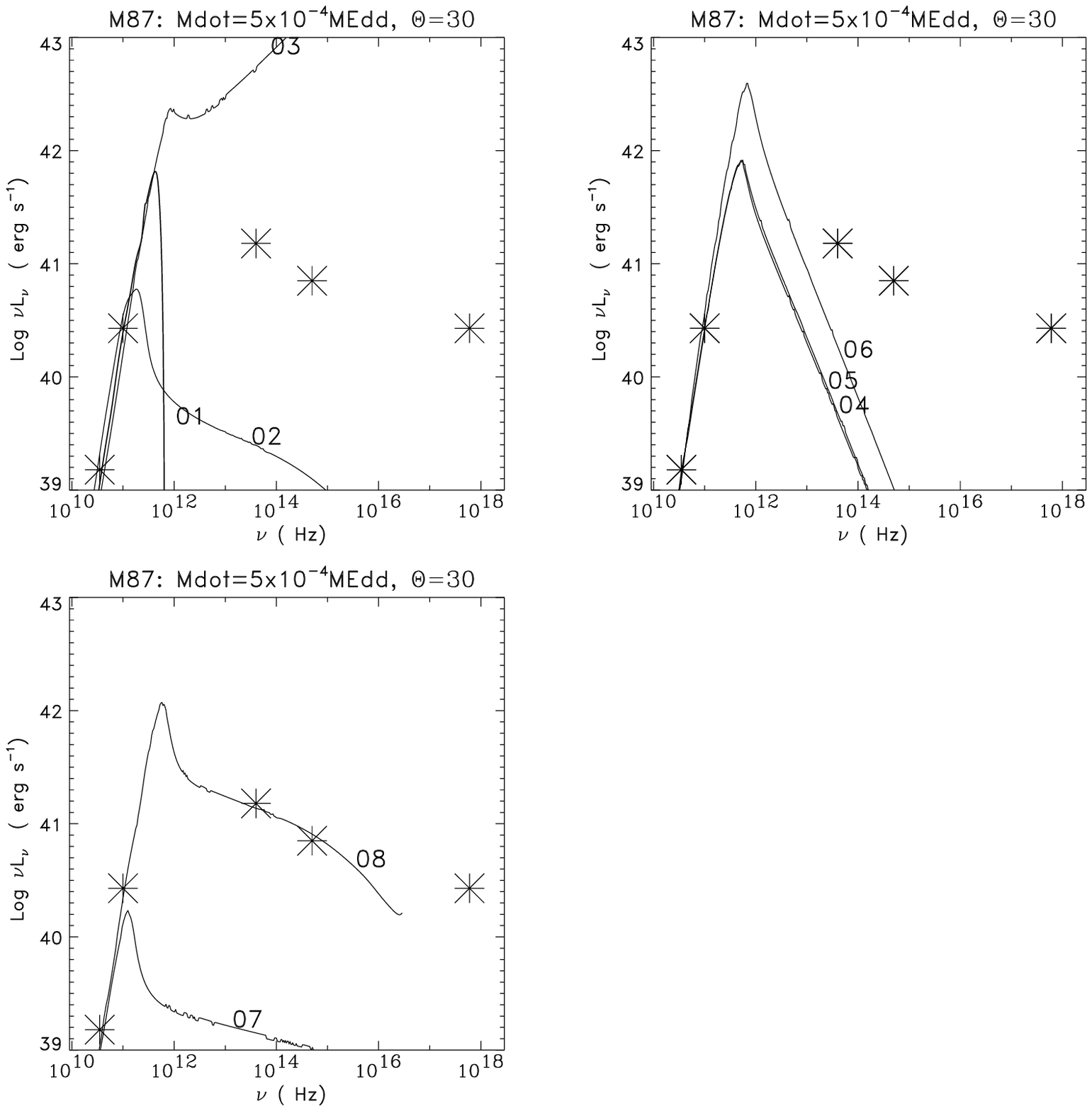}
}
\end{center}
{\vspace*{-0.4cm}}
\caption [ ] {  The SED of a BH-disk-jet model of the elliptical galaxy M87.
   In this model, an accretion rate of $5\times 10^{-4} \Mdot_\mathrm{Eddington}$
   and $T =5\times 10^6 $K
   is set to enter the domain of calculations through the outer boundary placed at 
    150 Schwarzschild radii from the central SMBH. The vertical scale hight of 
the disk at the outer
    boundary is taken to be $ H\approx 0.1 R_\mathrm{out}$. Additionally, a hot tenuous 
    corona is set to sandwich the optically thin disk.
   The calculated profiles (solid lines) are superposed on the observational 
   data (asterisks). The line 07 corresponds to a model in which  the PMF
is set to be in equipartition with the thermal energy, whilst  TMF=0.
 The line 08 is similar to 07, but the TMF is
 allowed to develop and reach values beyond equipartition with respect to the 
 thermal energy of the electron.
  } 
\end{figure}
 

\begin{figure}[htb]
\begin{center}
{\hspace*{-0.5cm}
\includegraphics*[width=7.5cm, bb=0 0 370 240,clip]
{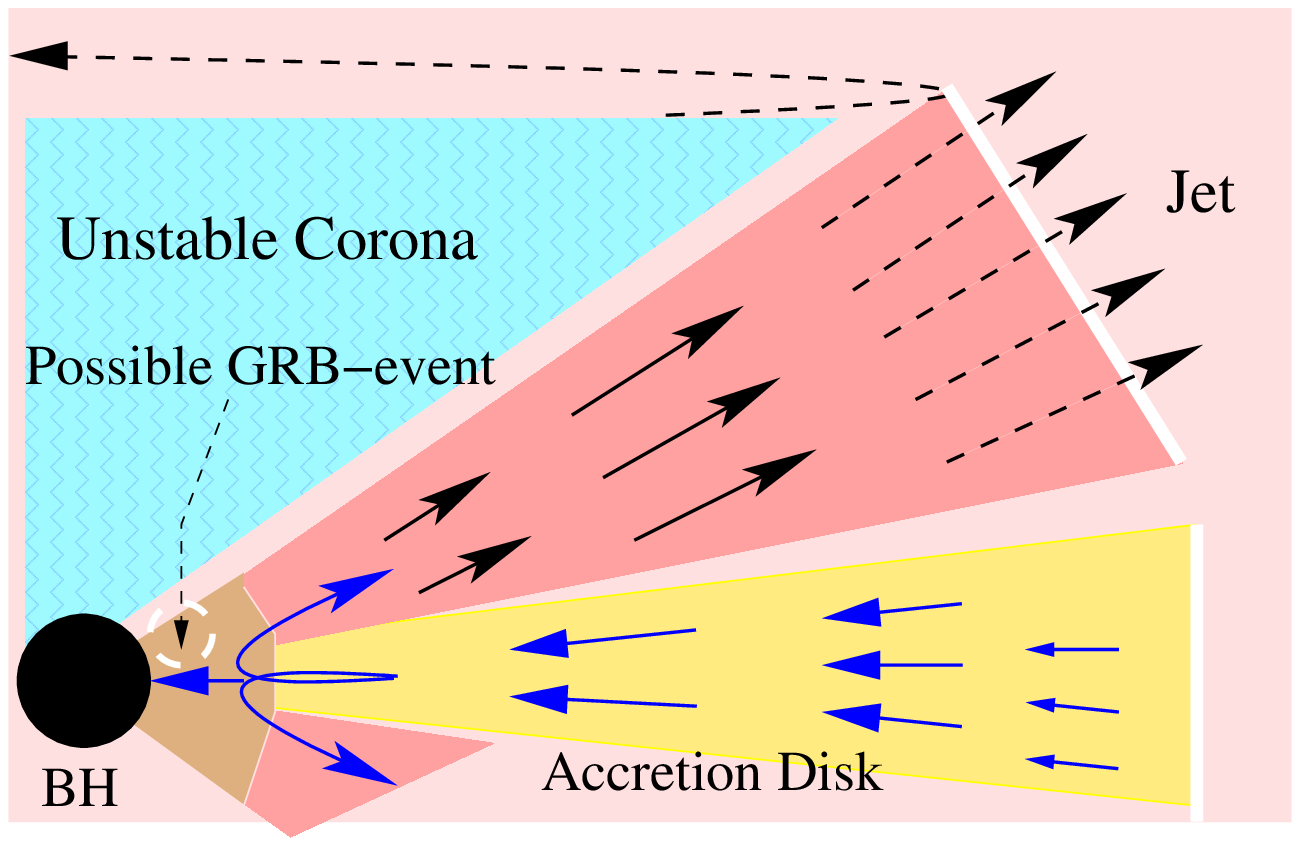}
}
\end{center}
{\vspace*{-0.4cm}}
\caption [ ] {  The matter in the disk is advection-dominated and flow inwards
with approximately the sound speed. Only in the innermost region, where a ion-dominated
torus is formed, matter start to move to higher latitude, while the associated
angular momentum force the matter to start centrifugal acceleration outwards. This
configuration places a key role for the central engine in powering the jet in the
vicinity of the event horizon. A spinning BH will certainly feed the jet with extra
power, whilst a non-rotating hole will extract energy from the jet.   } 
\end{figure}
 

\section{Formation of electron-proton jets: summary}

There are at least three ingredients that appears to be necessary 
 for initiating  jets from black hole
accreting systems:  
\begin{enumerate}
 \item large scale equipartition magnetic fields that are capable of extracting
     angular momentum from the disk efficiently, and deposit it in the viscous TL
      on the dynamical time scale,
 \item an underlying advection-dominated accretion disk for stably supplying 
   the plasma in the TL with angular momentum,
 \item strong shear that is capable of generating
  $B_\mathrm{T}$ beyond-equipartition with respect to the thermal energy of
the electrons on the dynamical time scale. The runaway amplification of $B_\mathrm{T}$ 
 is terminated then  by rapid reconnection in the magnetic-diffusive TL, 
thereby heating the proton-plasma up to the virial temperature. 
\end{enumerate}
This scenario appears to apply at least for the jet in the elliptical galaxy
M87 (see Biretta et al. 2002), where the central engine is believed to be a
 $3\times 10^9 \Msun$ supermassive
black hole (Fig. 5).
We note that since the flow in the TL is highly dissipative, we may expect
the flow in the TL to be causally connected with the central nucleus. 
This coupling may further strengthen the role of the central BH 
 in powering jet, 
therefore pronouncing possible observational differences between Kerr or Schwarzschild BHs..
\section{Formating beaming jets in GRBs}
Whether GRB-progenitors are dynamically unstable massive stars or mergers of
black holes with low mass compact stars (e.g., white dwarfs or neutron stars), 
the end-product is, most likely, a BH surrounded by a massive accretion disk
\cite{Shibata}, \cite{Baumgarte}. 
Unlike in AGNs and BHXBs, the accretion here proceeds on the dynamical time scale
and largely exceed the corresponding Eddington rates. Therefore, 
almost all the energies associated will be trapped in the accreted matter
 and disappear in the hole.
The shear in the disk excite torsional {\Alfven} waves that carry angular momentum
to higher latitudes, and whose speed of propagation is
 $V_\mathrm{A} = B_\mathrm{p}/\sqrt{\rho}$ (see Eq. 6). 
 The shear in the disk amplifies $B_\mathrm{T}$ in the first place, and subsequently 
 $B_\mathrm{p}$ via the BH-instability. During the amplification time 
$\tau_\mathrm{ampl.}$ ($= H_\mathrm{d}/V_\mathrm{A}$), both $B_\mathrm{p}$ and $B_\mathrm{T}$
depend weakly on the density.  This implies that 
\begin{equation}
\tau_\mathrm{rem}\simeq \tau_\mathrm{ampl.} \propto \rho \propto \Mdot .
\end{equation}
 Consequently,
the time scale for forming the TL where the jet-plasma rotate super-Keplerian, 
increases with increasing the accretion rate.  \\ 
On the other hand, since the time scale of angular momentum removal near the last stable orbit
is of the same order as the dynamical time scale, a non-negligible amount of
angular momentum will find its way still into the transition layer, thereby forming
a centrifugal barrier. The accumulated matter in this region, which becomes extremely
dense and optically thick even to neutrino emission, heats up via adiabatic compression,
shocks, and magnetic reconnection and forms a fermi torus in which the effective 
temperatures of the electrons,
protons, photons and neutrinos are of the same order of the virial 
{temperature}\footnote{This applies also when the hole is a Kerr with maximum rotation.}, i.e.,
\begin{equation}
 T_\mathrm{e} \simeq  T_\mathrm{p} \simeq  T_\mathrm{rad} \simeq  T_\mathrm{\nu}
\simeq  T_\mathrm{virial}. 
\end{equation}
The main cooling process here is though the loss of highly condensed matter 
that is being swallowed from below by the rapidly growing apparent horizon of 
the BH (see Fig. 6). In the case of a collapsar, the accretion rate prior to the burst
might be as large as $10^{14} \Mdot_\mathrm{Eddington}$ \cite{MacFadyen}.

Once the growth rate of the apparent horizon becomes comparable to, or longer than the
centrifugal acceleration rate, the accumulated super-Keplerian fermi layer starts its
runaway beamed explosion.

\def\Discussion{
\setlength{\parskip}{0.3cm}\setlength{\parindent}{0.0cm}
     \bigskip\bigskip      {\Large {\bf Discussion}} \bigskip}
\def\speaker#1{{\bf #1:}\ }
\def\endDiscussion{}

\Discussion

\speaker{Shibata}
How did you treat the corona? If the coronal density is low enough, 
the coronal MHD wind/jet would result. \\

\speaker{Hujeirat}   Yes. The corona is hot, tenuous
 and without rotational support. I would like to stress here that obtaining 
 coronal MHD wind, requires strong magnetic fields with an appropriate topology,
 and diffusive-free plasma. In our case, however,  the plasma in TL is highly
 diffusive, inducing a rapid reconnection of the TMF, and heating thereby the 
 plasma up to the virial temperature. It is because of the high magnetic diffusivity
 in the TL, torsional {\Alfven} waves are not able to transfer angular momentum
 into the dynamically-unstable corona on the dynamical time scale. 

\speaker{Fendt}
You said that you obtain a jet from your calculations. But they
do not look like that as they only show an uncollimated wind flow
along the disk surface. Please comment on that! \\

\speaker{Hujeirat} Jets are not jet-born, and collimation is a process that
 occurs on later times and on different length scales. What interests us here
 is not the jet-propagation and collimation, but the physical mechamisms underlying
 their formion and acceleration.
 What I have shown in this talk is that in the very early phases, jet-plasmas
 are outwards-accelerated via centrifugal forces. The TL appears to be the
 region where optimal acceleration is reached: torsional {Alfven} waves need to propagate
 a small length scale in the vertical direction to deposit the angular momentum 
 in the TL and accelerate the virial-hot protons outwards. This can be safely done
 on the dynamical time scale. Having the accelerated low-cooling protons crossed 
 a critical radius, beyond which the magnetic diffusivity becomes negligibly small, the TMF
 will take care then to re-direct and collimate the flow into jets. It should be
 noted here  that the ion-dominated and super-Keplerian rotating plasma in the TL
 has a large positive energy, hence gravitationally unbound and potentially could
 expand to infinity.
 
\endDiscussion
 
\end{document}